\documentclass[12pt]{article}\usepackage[hyperfootnotes=false]{hyperref}
   \usepackage{epsfig}
      \usepackage{amsmath}
      \usepackage{amssymb}
  \usepackage{graphicx}
  \newlength{\abstractwidth}
  \renewcommand{\thefootnote}{\fnsymbol{footnote}}
  \renewcommand{\thanks}[1]{\footnote{#1}} % Use this for footnotes
  \newcommand{\starttext}{
  \setcounter{footnote}{0}
  \renewcommand{\thefootnote}{\arabic{footnote}}}
  \renewcommand{\theequation}{\thesection.\arabic{equation}}
  \newcommand{\be}{\begin{equation}}
  \newcommand{\bea}{\begin{eqnarray}}
  \newcommand{\eea}{\end{eqnarray}}
  \newcommand{\beq}{\begin{equation}}
  \newcommand{\ee}{\end{equation}}
  \newcommand{\eeq}{\end{equation}}

  \def\ba{\begin{eqnarray}}
  \def\ea{\end{eqnarray}}

  \def\12{{1 \over 2}}
  \def\eq{&=&}

  \def\d{\partial}

  \def\cc{cosmological constant }
  
  \def\simleq{\; \raise0.3ex\hbox{$<$\kern-0.75em
      \raise-1.1ex\hbox{$\sim$}}\; }
   \def\simgeq{\; \raise0.3ex\hbox{$>$\kern-0.75em
      \raise-1.1ex\hbox{$\sim$}}\; }

\def\cdl{Coleman De Luccia}

\def\ba{\bf{a}}

  \def\h3{{\cal{H}}_3}

\def\o3{\Omega_3}

\def\O2{\Omega_2}
\def\o{\omega}
 \def\d2{$dS_{2+1}$}

 \def\bi{\begin{itemize}}
  \def\ei{\end{itemize}}

      \def\2{   $2$-adic}

\begin{document}
  \renewcommand{\theequation}{\thesection.\arabic{equation}}
  \begin{titlepage}
  \rightline{}
  \bigskip

  \bigskip\bigskip\bigskip\bigskip

    \centerline{\Large \bf {Fractal-Flows and Time's Arrow}}
    \bigskip

  \bigskip \bigskip

  \bigskip\bigskip
  \bigskip\bigskip
  %\centerline{\it }
  %\medskip
  %\centerline{} \centerline{} \centerline{}
  %\bigskip

  \begin{center}
  {{ Leonard Susskind}}
  \bigskip

\bigskip
Stanford Institute for Theoretical Physics and  Department of Physics, Stanford University\\
Stanford, CA 94305-4060, USA \\

\vspace{2cm}
  \end{center}

  %\bigskip\bigskip

 %\bigskip\bigskip
  \begin{abstract}

% Abstract

This is the written version of a lecture at the KITP  workshop on Bits, Branes, and Black Holes.
In it I describe work  with D. Harlow, S. Shenker,  D. Stanford  which explains how the  tree-like structure of eternal inflation, together with the existence of terminal vacua, leads to  an arrow-of-time.   Conformal symmetry of the dS/CFT type is inconsistent with an arrow-of-time  and must be broken.  The presence in the landscape of terminal vacua leads to a new kind of attractor called a fractal-flow, which both breaks conformal symmetry, and creates a directional time-asymmetry. This can be seen from both the local or causal-patch viewpoint, and also from the global or multiversal viewpoint. The resulting picture is consistent with the view recently expressed by Bousso.

In the last part of the lecture  I illustrate  how the tree-model can be useful in explaining the value of the cosmological constant, and the cosmic coincidence problem. The mechanisms are not new but the description is.

 \medskip
  \noindent
  \end{abstract}

  \end{titlepage}
  \starttext \baselineskip=17.63pt \setcounter{footnote}{0}

%%%%%%%%%%%%%%%%
%%%%%%%%%%%%%%%%
%%%%%%%%%%%%%%%%
%%%%%%%%%%%%%%%%
%%%%%%%%%%%%%%%%
%%%%%%%%%%%%%%%%

\tableofcontents

\setcounter{equation}{0}
\section{Introduction}

\subsection{Big Numbers}

Of the things I am going to discuss, the most difficult to accept may be the incredibly large numbers that are an unavoidable part of them.  We have already had to swallow some  pretty big numbers.  For example, the observed flatness of the universe implies that it is much bigger than the horizon scale---by how much no one knows.

Again: the  landscape of vacua may be  orders of magnitude larger---even in the logarithm---than the often-quoted $10^{500}$. Again, no one knows.

 What has been less appreciated is the  time-scales over which the mechanisms of eternal inflation take place. It will take a few hundred billion years for the CMB  to thermalize to the local de Sitter temperature, but for the large-scale attractors that eternal inflation  envisions, the time scale for equilibration is at least $e^{10^{120}}$.  Skepticism about such monstrous time intervals is perfectly legitimate.  But in their defense I would make two points. The first is that these \it are \rm the time-scales of eternal inflation: to discuss eternal inflation but to reject their legitimacy  is in my opinion illogical.

The second remark is that eternal inflation may have explanatory power.  (In the last part of this lecture I will review some examples.)    It offers explanations for the small value of the cosmological constant; the strange coincidence between the time at which observations are made, and the onset of cosmic-acceleration;   and as we will soon see, the arrow of time. But all of these explanations require us to consider the statistics of super-huge distances, times, and landscapes. So I ask you to keep an open mind.

\subsection{The Boltzmann-Fluctuation Problem}

It has been claimed that the  arrow-of-time can only be explained by assuming that the initial entropy of the universe is very small. In an eternally inflating universe this is not correct; a low entropy initial condition is neither necessary nor sufficient for an  arrow-of-time.  Let's begin with whether it is sufficient. Suppose there is a  theory that contains only vacua with positive cosmological constants, and some agent prepares the universe  in a state of very low entropy. For example, the state of lowest entropy would be the vacuum with largest cosmological constant.

The initial vacuum would decay, through a chain of tunneling events, to the state of smallest cosmological constant. Each decay would be followed by a transient period during which an  arrow-of-time would exist, until the vacuum reached de Sitter thermal equilibrium. During such a transient period observers would see a normal directional flow of time.

The problem with this theory, when viewed from the super-long time-scales of eternal inflation, was pointed out by Dyson, Kleban, and Susskind \cite{Dyson:2002pf}. Once  maximum entropy is reached, there is an infinite amount of time for ergodicity to create large thermal fluctuations with \it local  \rm uncorrelated arrows of time.
\begin{figure}[ht]
\begin{center}
\includegraphics[scale=.3]{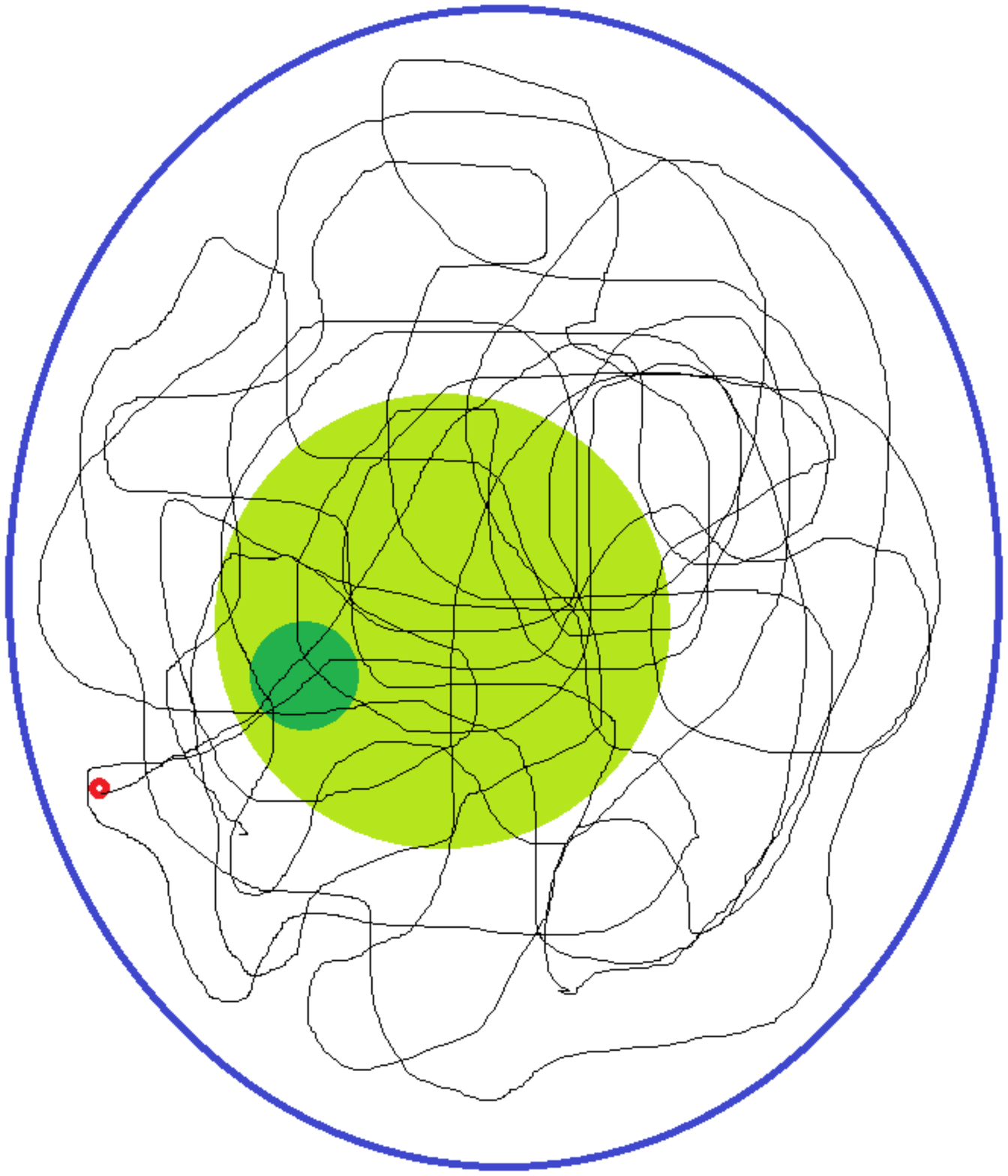}
\caption{An ergodic path through the phase space of an inflating universe. The small red dot is a very low-entropy starting point such as an inflating geometry with a large vacuum energy. The dark green disk is the region which looks like our universe. The light green region contains observer-friendly  freak-universes. The overwhelming number of trajectories that pass into either of the green areas do not flow directly from the tiny red dot. A trajectory will flow from the dark green disk into the red dot as frequently as it flows out of it. }
\label{f6}
\end{center}
\end{figure}

These Boltzmann fluctuations permit structure to form and evolve, but the overwhelming number of such fluctuations are freak occurrences which look nothing like our world. Freak-worlds would  statistically dominate, implying that a ``normal" universe would be the extraordinary exception. If time is viewed as a river, then this kind of theory would describe a damned up channel in which the only flows are local thermal fluctuations. What we need is a mechanism to create a global flow.

If a low-entropy initial condition is not  sufficient  for an  arrow-of-time, is it  necessary? The answer is no: as Bousso has argued, ``up-transitions" (fluctuations in which the cosmological constant increases) can allow a causal patch to jump to a low entropy state from a larger entropy state.  From there it may evolve normally \cite{Bousso:2011aa}. As Bousso himself argues,  this can only be viable if, in addition to de Sitter vacua, the landscape contains  so-called terminal vacua. Terminal vacua are causal patches that have exited from the prevailing inflation by having either zero or negative cosmological constant.
Basically, the dominance by Boltzmann fluctuations is avoided if the vacuum exits inflation well before the recurrence time.

As I will explain, the fractal-flow is a new mathematical attractor that replaces conformal invariance when terminals are present, and induces an  eternal arrow-of-time\footnote{Professor Nomura has kindly informed me of similar ideas in his paper \cite{Nomura:2011dt}}.

\subsection{Eternal de Sitter Space?}

 Let me dispel a false idea. Suppose there is a landscape described by a collection of scalar fields, and a number of strictly positive local  minima of the potential. One normally thinks of each local minimum as a de Sitter vacuum, but strictly speaking there is only one equilibrium state. Either from a local or a global perspective, the vacua make tunneling transitions---both down and up---so that an eventual equilibrium is set up. That equilibrium would be the only exact de Sitter vacuum.

From a causal patch viewpoint this is just thermal equilibrium in which up and down-transitions equilibrate. In  equilibrium  the probability for any given metastable vacuum $m$  is given by
\be
P(m) = z^{-1} e^{S_m}
\ee
where $z$ is a normalization factor and $S_m$ is the usual Hawking Gibbons entropy of the local minimum. Put another way, all microstates are equally probable.

From the global or multiversal viewpoint the same arrow-less equilibrium is described by a  conformally invariant fixed-point of a putative dS/CFT  \cite{Strominger:2001pn}.  The lack of an arrow-of-time makes conformal invariance  a thing to be avoided.

This raises a  question: Is a theory of absolutely stable de Sitter equilibrium  mathematically consistent?   The answer may be no; the non-compact symmetries of de Sitter space---conformal symmetries in dS/CFT---are inconsistent with the finite entropy of a causal patch. The no-go theorem was proved ten years ago by Goheer,  Kleban, and Susskind \cite{Goheer:2002vf}. I view this as a positive thing; it mathematically rules out  a class of theories which are phenomenologically unviable. ( The  theorem of \cite{Goheer:2002vf} makes explicit reference to recurrence time-scales. Presumably, approximate theories of de Sitter space can make sense for ordinary cosmological times.)

The purpose of this lecture is explain, in as mathematically a precise fashion as I can, the relations between eternal inflation, conformal  fixed-points, terminal vacua, and the arrow-of-time. To do so I will rely on a model of the tree-like structure of eternal inflation   \cite{Harlow:2011az}. What one finds in the model, and I believe it to be true more generally,
 is that introduction of terminal vacua leads to a new kind of attractor-behavior called  a \it fractal-flow. \rm   The  fractal-flow has a robust arrow-of-time, not dependent on fluctuations.  It has a  universal scaling behavior but it is not a CFT.  In fact it is not a field theory in any ordinary sense.

\setcounter{equation}{0}
\section{The Tree Model Without Terminals}

I will begin with a short review of the Harlow, Shenker, Stanford, Susskind model.
The causal structure of the model multiverse is a tree-graph that grows out of a root in the remote past. At each node of the tree, one incoming edge splits into some number, $p$ of outgoing edges, sometimes called cells in \cite{Harlow:2011az}. The simplest case of $p=2$ is sufficient for my purposes and I will adopt it from now on. Each doubling of the number of edges will be called a 2-folding.

\begin{figure}[ht]
\begin{center}
\includegraphics[scale=.3]{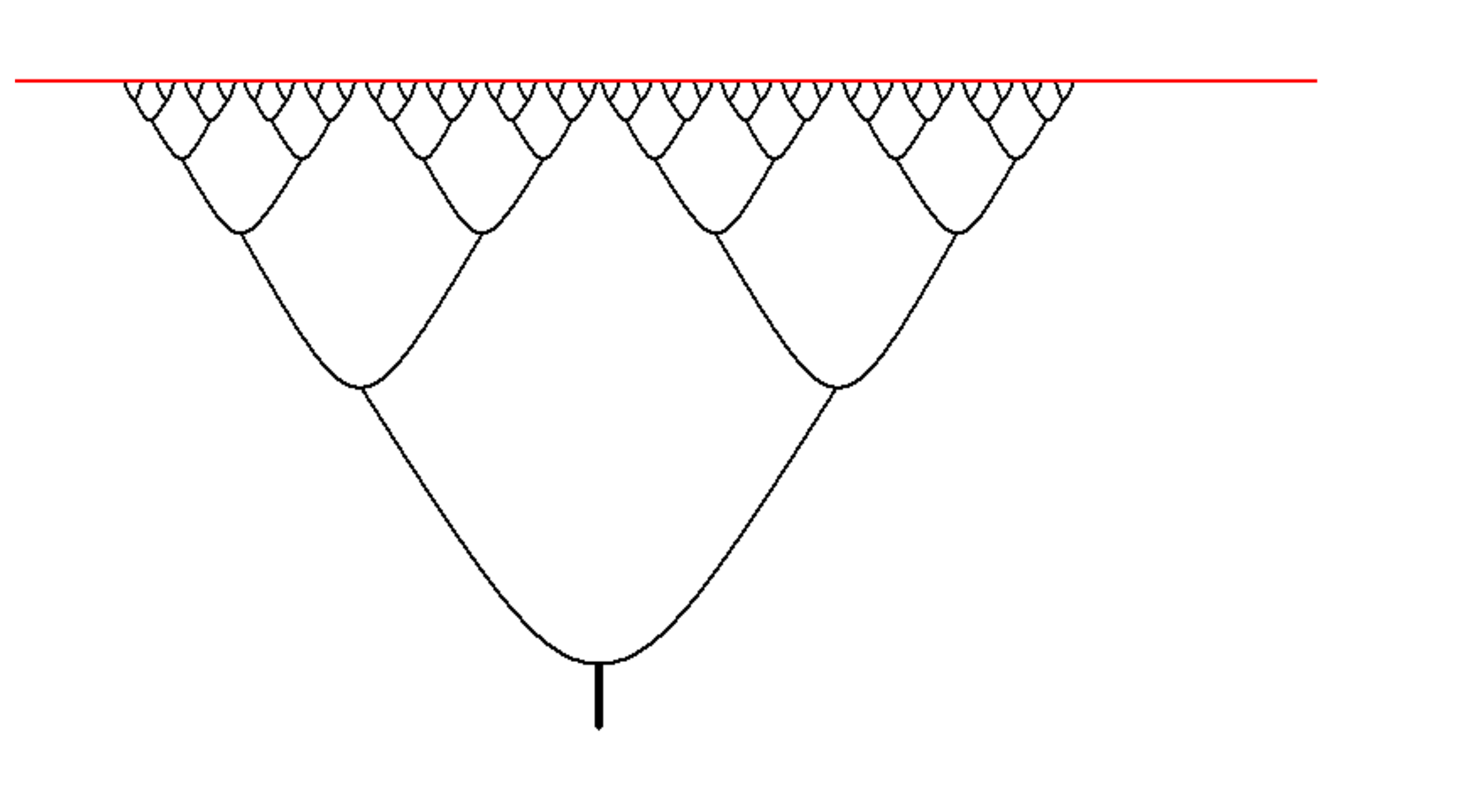}
\caption{The eternally inflating tree and its future boundary.}
\label{f6}
\end{center}
\end{figure}

The two concepts of \it  future boundary \rm and  \it causal patch \rm are central to the tree model. Let us define them.
The tree has a future asymptotic boundary which is the precise analog of the future boundary of de Sitter space.  The  future boundary of a tree is not  smooth geometry in the usual sense. One defines it by following a future-directed path to $u= \infty.$  Each such  path ends on a point $x$  of the future boundary. As explained in  \cite{Harlow:2011az} the set of such points defines the \2 \ number system.  Go to any point $x$ on the  \2 \  boundary and run backward along a unique sequence of edges to the root. That sequence of edges is the causal patch of the future boundary-point $x.$

The tree exponentially expands so that after $u$  2-foldings the number of nodes and edges is $2^u.$  Each edge of the tree has a ``color" $m$ which represents a vacuum on a strictly positive \cc \ landscape. It is assumed that the color $m$ represents a collection of microstates described by an entropy $S_m.$

At each node the incoming edge with vacuum-color  $n$ splits into two edges. Each new edge has probability $\gamma_{nm}$ to make a transition from color $m$ to color $n.$ The transition matrix $\gamma_{nm}$ plays the role of the dimensionless tunneling rate for a transition from vacuum type $m$ to type $n.$ The fundamental property of microscopic reversibility is implemented by the condition of detailed balance,
\be
\frac{\gamma_{nm}}{\gamma_{mn}} = e^{S_n-S_m}.
\label{detail}
\ee

Equivalently, detailed balance can be expressed in terms of a real symmetric matrix $M_{nm} = M_{mn}:$
\be
\gamma_{nm} = M_{nm} e^{S_n.}
\label{symmetric M}
\ee

 Let us define the probabilities $P_m(u)$  as the fraction of edges with color $m$ after $u$  2-foldings of the tree.
 The  $P_m$  satisfy the rate equations
 \be
 P(u+1)  = G P(u)
 \label{rate}
 \ee
where $P$ is a column-vector composed of the $P_m;$  and $G$  is the matrix,
\be
G_{mn} = \delta_{mn} - \sum_r \gamma_{rm}   \delta_{mn}  + \gamma_{mn}
\label{G-matrix}
\ee

The matrix $G$ is not symmetric so its eigenvectors are not orthogonal, but detailed balance allows us to transform the equation to a form in which the eigenvectors are orthonormal. To that end, define the diagonal matrix
\be
Z_{mn}=\delta_{mn} e^{S_n/2}.
\ee
The matrix
\be
S=Z^{-1} G Z
\ee
is seen to be symmetric. The column vectors can also be transformed:
\be
\Phi = Z^{-1} P.
\ee
The rate equation in terms of $\Phi$ has the symmetric form,
\be
\Phi(u+1) = S \Phi(u).
\ee
Note that the original probabilities $P_m$  are given by
\be
P_m= e^{S_m/2}\Phi_m
\label{p=es phi}
\ee

Since $S$ is symmetric its eigenvectors are orthogonal. We denote them by the symbols  $(I)_m.$  The label $I$ runs over the same number of values as does the color label $m.$  The $(I)$ satisfy the eigen-equations
\be
S \cdot (I) = \lambda_I (I)
\ee

The eigenvectors satisfy
\be
(I)\cdot (J) = \sum_m (I)_m(J)_m  = \delta_{IJ.}
\ee
and
\be
\sum_I (I)_m (I)_n = \delta_{mn}
\ee

\subsection{Equilibrium}

All of the eigenvalues $\lambda_I$  are positive. The largest eigenvalue is equal to $1$ and all others are less than $1.$  We will write the eigenvalues in terms of  scaling-dimensions\footnote{I will justify the term scaling-dimension shortly.}  $\Delta_I,$
\be
\lambda_I = 2^{-\Delta_I}
\ee
Note that the leading eigenvalue $\lambda_0 = 1$ corresponds to scaling dimension zero. Typically the other dimensions are very small, being of order the tunneling rates $\gamma.$

The eigenvector corresponding to eigenvalue $1$ is called $(0)$ and its elements are given by
\be
(0)_m \sim e^{S_m/2}.
\ee

The first thing to notice about $(0)$ is that it is the only eigenvector which does not tend to zero with the time $u.$ All others fade to zero like
$2^{-\Delta_I u}$ Thus $(0)$ represents  the equilibrium distribution of the colors, and the other eigenvectors represent transients.

 In terms of the probabilities $P$  the equilibrium distribution is
\be
P_m^{(0)} =z^{-1} e^{S_m  }
\label{equilibrium}
\ee
where $z^{-1}$  is a normalization factor.

There are two meanings to the probabilities $P_m^{(0)}:$  one global, one local. Globally $P_m^{(0)}$ is the fraction of edges that have color $m$ after transients have faded away. They are completely independent of the initial conditions. At a given time $u$ the total number of edges is
\be
N_{total} = 2^u
\ee
and the number of edges of type $m$ is
\be
N_m= 2^u P_m(u) \to z^{-1} e^{S_m  } 2^u .
\ee

The other interpretation of $P$ is local. As we follow a causal patch from the root of the tree to the future boundary, the color of the edges will vary according to the rate equation. The quantity $P_m(u)$ is the probability that at time $u$ the color is $m.$  After initial transients fade away, the probability that the causal patch has color $m$ is $P_m^{(0)}.$  This interplay of local and global is called local/global duality \cite{Bousso:2009mw}.

One final point is that the time scale for the $P_n$ to fully equilibrate to the  distribution \ref{equilibrium}, from an arbitrary initial condition, is determined by the slowest rates in the rate equation. These will be the up-transitions. For the up-transition from $n$ (a state of large entropy) to $m$ (a state of low entropy) the rate is of order $e^{-S_n}.$ Thus the time for full equilibration is of order $e^{S}$ where $S$ is the maximum entropy on the  de Sitter landscape. That is why I said the time scales associated with eternal inflation are at least as big as $e^{10^{120}}.$  The number $10^{120}$ is the entropy of our local de Sitter vacuum.

\subsection{Global Correlations}

To study global correlations we temporarily cut off the time at some large time $u =u_0,$ thereby defining a regulated version of the future boundary. The points of the regulated boundary are called $x.$  There is a natural metric on the (regulated) boundary defined in the following way. Pick any two points $x, \ y .$  From each point there is a unique path of edges that runs backward to the root. Run backward along the paths from $x$ and $y$  until they intersect at a node, and define the time of intersection to be $u_i.$   The  \2 \ distance between $x$ and $y$  is defined by
\be
|x-y|_2 \ = \ {2^{-u_i}}
\label{distance}
\ee
The \2 \ distance is a measure of how far into the past you have to follow the causal patches until they come into causal contact. Note that the distance defined in this way is not sensitive to the cutoff $u_0.$

A system of correlation functions at  time $u_0$  can be defined as follows. The one-point function $C_n(x)$ is defined as the probability that the edge at $x$  has color $n.$  Assuming that $u_0$ is late enough that transients have died out, $C_n(x)$  will be independent of both $u_0$  and $x.$  It is obviously given by the equilibrium distribution,
\be
C_n(x) = P_n^{(0)} = \frac{1}{z}e^{S_n}
\ee

Next consider the correlation function $C_{nm}(x,y),$  defined as the joint probability that the edges leading into $x$ and $y$  have colors $n$  and $m.$  The  only source of correlation between $x$ and $y$ is the fact that  they trace back to the common node at time $u_0.$  Suppose the color of the edge coming into that node is $r.$

Define $P_{nr}(u)$  to be a conditional probability that a causal patch  at time $u_i+u$  will have color $n,$  given that at time $u_i$ it has color $r.$   In \cite{Harlow:2011az}  $P_{nr}(u)$   was called a propagator, although it has nothing to do with field theoretic propagators.
\begin{figure}[ht]
\begin{center}
\includegraphics[scale=.3]{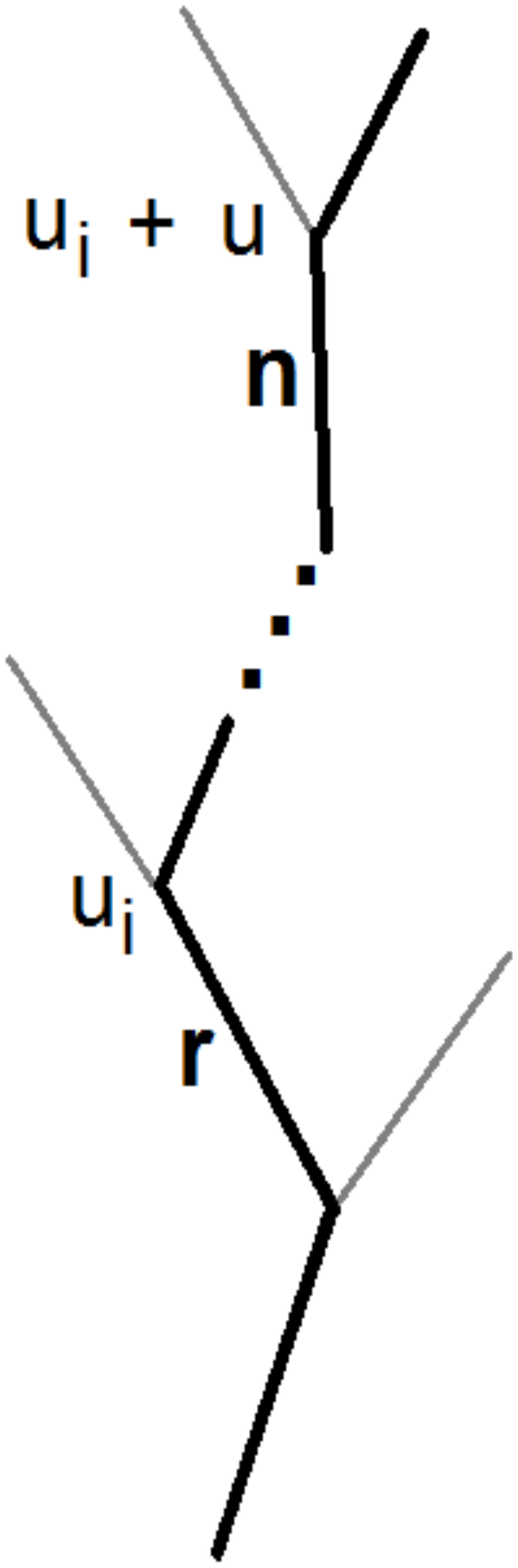}
\caption{The propagator is the probability that a causal patch (the dark sequence of edges)
at time $u_i+u$  will have color $n,$  given that at time $u_i$ it has color $r.$.}
\label{f6}
\end{center}
\end{figure}

 The propagator is
 given by
 the $u$-th power of the matrix $G$:
\be
P_{nr}(u) \equiv (G^u)_{nr}= (I)_n \ \lambda_I^u \ (I)_r \ e^{\frac{S_n-S_r}{2}}.
\label{connector}
\ee

Since once the branches leading to $x$  and $y$ split off from one another they have no further interaction, the
 probability that $x$ and $y$ have color $n$ and $m$ has the form
\be
P_{n r}(u_0 -u_i)P_{m r}(u_0 -u_i).
\label{partial prob}
\ee

To complete the computation we multiply \ref{partial prob}  by the probability that the common node at $u_i$
has color $r,$ and then sum over $r.$
\be
C_{nm}(x,y)=\sum_r P_r  P_{n r}(u_0 -u_i)P_{m r}(u_0 -u_i).
\label{C=PPP}
\ee
If we assume that the time of the intersection, $u_i$ is late enough that transients have died away, then we can replace $P_r$ by the equilibrium eigenvector and we find
\be
C_{nm}(x,y)=\frac{1}{z}\sum_r e^{S_r}  P_{n r}(u_0 -u_i)P_{m r}(u_0 -u_i).
\label{C=ePP}
\ee

The final step is to substitute \ref{connector} and use the orthonormality of the eigenvectors $(I)$ to obtain,
\be
C_{nm}(x,y)= \frac{1}{z}\sum_{I}
\lambda_I^{2 (u_0-u_i)} \ (I)_{n} \ (I)_{m} \ e^{S_{n}+S_{nm} \over 2}.
\ee
or, using \ref{distance} we find the suggestive form,
\be
C_{nm}(x,y)= \frac{1}{z}\sum_{I}
\lambda_I^{2 u_0} \ (I)_{n} \ (I)_{m} \ e^{S_{n}+S_{nm} \over 2}   \frac{1}{|x-y|_2^{\Delta_I}}.
\label{suggestive}
\ee

The interpretation is obvious. The correlation function is a superposition of conformal correlation functions for fields of dimension $\Delta_I.$ The pre-factors $\lambda_I^{ u_0}$ for each external line are wavefunction renormalization factors. Stripping off these factors we find a system of conformal correlation functions $\langle  {\cal{O}}_I(x)   {\cal{O}}_J(y)     \rangle$  that satisfy
\be
\langle  {\cal{O}}_I(x)   {\cal{O}}_J(y)     \rangle = \frac{\delta_{IJ}}{|x-y|_2^{2\Delta_I}}
\ee

There are a number of interesting things about this result. First of all, consider the term in \ref{suggestive} associated with the leading eigenvector $(I) = (0).$ Since $\Delta_0=0$ this term is given by the factorized form,
\bea
C_{nm}(x,y)\eq \frac{1}{z}
  (0)_{n} \ (0)_{m} \ e^{S_{n}+S_{nm} \over 2}   \cr \cr
  \eq P_n^{(0)}P_m^{(0)}
\label{cluster}
\eea

Since the other terms tend to zero as inverse powers of  $|x-y|_2$   the two-point function satisfies cluster decomposition.
Remarkably the conformal correlation functions  have the precise form of CFT correlation functions for fields of dimension $\Delta_I $  except that the usual distance $|x-y|$  is replaced by the \2 \ distance $|x-y|_2.$

This correspondence with conformal field theory is not accidental. As was shown in  \cite{Harlow:2011az}, there is an underlying symmetry of tree-graphs that acts on the  boundary as \2 \ conformal transformations.

However---and this is the important point---the symmetry only acts on graphs made of undirected edges. The edges of the tree-model are supposed to be future-directed and in general the directed-ness destroys the \2 \ conformal symmetry. But as long as detailed balance is satisfied the symmetry is restored and the correlations are those of a CFT on the \2 \ boundary.  The similarity with CFT correlations is not restricted to two-point functions and includes such things as operator product expansions.
For example, the three point function has the usual conformal form
  \be
  \langle \mathcal{O}_I(x)\mathcal{O}_J(y)\mathcal{O}_K(z)\rangle
=
 \frac{C_{IJK}}{|x-y|_p^{\,\Delta_I+\Delta_J - \Delta_K} |x-z|_p^{\,\Delta_I+\Delta_K-\Delta_J} |z-y|_p^{\,\Delta_K + \Delta_J - \Delta_I}}.
 \label{3pt1}
  \ee
  with
  \be
C_{IJK} = \mathcal{N}^{1/2}\sum_n e^{-S_n \over 2} \ (I)_n \ (J)_n \ (K)_n.
 \label{3pt2}
 \ee

The structure constants $C_{IJK}$ satisfy the associativity condition,
\be
\sum_H  \ C_{IJH}C_{HKL} =\sum_H  \ C_{IKH}C_{HJL}
\ee

\subsection{Multiverse Fields}

The fields ${\cal{O}}_I(x) $ are not any of the usual perturbative   fields such as string-theory moduli. They are defined as transforms of fields ${\Phi}_n(x) $ which take on the value $1$ at a point in the multiverse where the color is $n$ and zero elsewhere. The ${\cal{O}}_I(x) $   are fields of definite dimension, whereas the ${\Phi}_n(x) $ are linear superposition with different dimensions.  The ${\Phi}_n(x) $ and the ${\cal{O}}_I(x) $  are non-perturbative objects that only make sense on scales at least as big as a causal patch---in other words super-horizon scales. The purpose of these \it multiverse fields \rm is to study global properties of the multiverse.

The characteristic feature of multiverse fields is their extremely low dimensionality. The dimensions $\Delta_I$ are generally proportional to the transition rates $\gamma_n$  which in a realistic theory would be \cdl \ tunneling rates. Moreover, if the landscape is as rich as string theory suggests \cite{Bousso:2004fc}\cite{Kachru:2003aw}\cite{Susskind:2003kw} , then there should be a huge collection  of such fields and a discretuum of operator dimensions.

It is clear that the non-perturbative version of a de Sitter---conformal field theory correspondence would have to contain this kind of vast array of multiverse fields. In particular in the doubled-dS/CFT version of \cite{Harlow:2012dd}  they would correspond to source fields (arguments of the Wheeler de Witt functional) which are integrated over in calculating expectation values.

\subsection{Causal Patch Correlations}

From an observational point of view we are not interested in the global view but only the view from a causal patch \cite{Bousso:2009mw}. Causal patches are of course embedded in a growing tree. It is not completely obvious whether or not the causal patch inherits an arrow-of-time from the surrounding expansion. To see that it does not we can directly investigate correlations within a causal patch.

Consider the following question: Given that a point on a causal patch has color $m$ at time $u,$ what is the probability that it will have color $n$  at the next instant, $u+1?$  I'll write this probability as
$$P(n, u+1 |  m,u  ).$$

The answer is obviously an element of the rate matrix $\gamma_{nm}.$  Using \ref{symmetric M} we can write this in the form
\be
P(n, u+1 |  m,u  ) = M_{nm} e^{S_n}.
\ee

Now let us ask the time-reversed question:  Given that a point on a causal patch has color $m$ at time $u,$ what is the probability that it had color $n$  at the previous instant $u-1?$  From Bayes' theorem,
\be
P(n, u-1 |  m,u  ) =   P(m,u  |   n, u-1   )  \frac{P(n, u-1)}{P(m,u )}.
\label{bayes}
\ee

The probability $P(m,u  |   n, u-1   )$  is the transition matrix element $\gamma_{mn}.$ This gives,
\be
P(n, u-1 |  m,u  ) =  \gamma_{mn}  \frac{P(n, u-1)}{P(m,u )}.
\ee

In general this is nothing special, but if the transients have died out by time $u$  we may write this as
\be
P(n, u-1 |  m,u  ) =  \gamma_{mn} e^{S_n-S_m}
\ee
or
\be
P(n, u-1 |  m,u  ) =  M_{mn} e^{S_n}
\ee

Given the symmetry of $M$ we find
  \be
P(n, u-1 |  m,u  ) =  P(n, u+1 |  m,u  )
\ee

In other words there is a complete time-reversal symmetry and a total absence of an  arrow-of-time.  We could easily extend this to more general situations---for example,
  \be
P(n, u-v |  m,u  ) =  P(n, u+v |  m,u  )
\ee
for any integer  $v.$

The message to be taken away from this is that conformal symmetry of the future boundary of an eternally inflating de Sitter geometry is associated with the same  time-reversibility ( detailed balance)  that leads to  thermal equilibrium  in a causal patch.
It follows that a theory with an  arrow-of-time must not have a conformally invariant future boundary.

\setcounter{equation}{0}
\section{The Tree Model With Terminals}

The existence of terminal vacua changes the
story in a very positive direction,
by a mechanism that is simply illustrated in the
tree model. With terminals, the conformally invariant fixed-point is
replaced by a new kind of attractor---one  that does have an arrow-of-time.

In the presence of terminal vacua, the co-moving coordinate volume (but not the proper volume) eventually becomes
solidly dominated by terminals. Eternal inflation does not end, but it is restricted to a
fractal of coordinate volume that decreases like $\lambda_D^u$
 ($\lambda_D^u$ being the dominant eigenvalue, soon to be defined).

 Despite the fact that the coordinate volume goes to zero, it is
still possible to define a collection of non-trivial conditional correlation functions on future
infinity:
Following \cite{Harlow:2011az} we add terminal vacua to the tree model. This is done by randomly pruning the tree.
\begin{figure}[ht]
\begin{center}
\includegraphics[scale=.3]{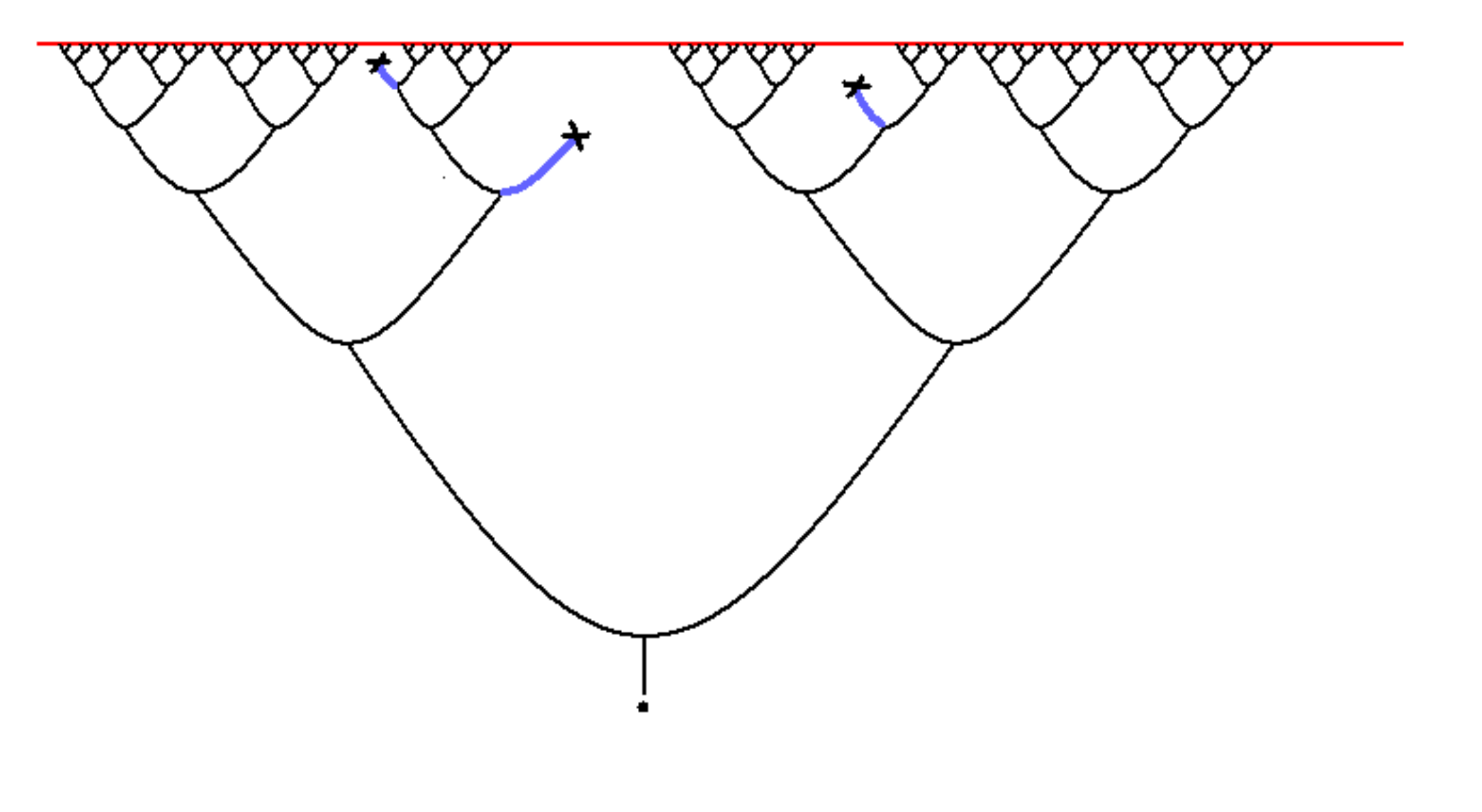}
\caption{The pruned tree of a landscape with terminals.}
\label{f6}
\end{center}
\end{figure}
Suppose an edge has color $n.$  When it branches the probability that  a daughter edge is terminal will be denoted $\gamma_n.$  The rate equations for non-terminal vacua are still a closed set of equations. Equations \ref{detail}, \ref{symmetric M}, and \ref{rate}, are unchanged but the expression \ref{G-matrix} for the rate matrix is replaced by,

\be
G_{mn} = \delta_{mn}(1-\gamma_n) - \sum_r \gamma_{rm}   \delta_{mn}  + \gamma_{mn}
\label{G'-matrix}
\ee
These equations no longer satisfy probability conservation for the simple reason that probability leaks into the terminal vacuum.

The matrix $S=Z^{-1}GZ$ is still symmetric and its eigenvectors still form an orthonormal basis, now with all eigenvalues less than one in magnitude.
The largest eigenvalue---the so-called dominant eigenvalue---of this  modified $G$-matrix determines the asymptotic late-time population of non-terminal vacua.  The dominant eigenvalue also defines a dominant scaling dimension,  $2^{-\Delta_D} = \lambda_D.$

The corresponding eigenvector, denoted $(D)_m,$  is called the dominant eigenvector.
 We also define the probability $P^{\{D\}}_m$  through  a version of equation \ref{p=es phi}
 \be
 P^{\{D\}}_m = e^{\frac{S_m}{2}} (D)_m.
 \ee

The  probabilities  $P^{\{D\}}_m$  are no longer proportional to $e^{S_n}.$  They tend to be heavily weighted toward the longest lived vacua and not necessarily the largest entropy vacua.
It should be noted that although the dominant eigenvector has time dependence
\be
2^{-\Delta_D u}
\ee
the decrease in its magnitude does not mean that eternal inflation ends. The volume of inflating vacuum grows a little slower than previously, with the time dependence
\be
2^{(1-\Delta_D) u}
\ee
However, the probability for survival along a trajectory tends to zero.

As in the case without terminals, propagators can be defined by \ref{connector} but their actual values are determined by the matrix
\ref{G'-matrix} and not \ref{G-matrix}.

\subsection{ Fractal-Flow Behavior}

I have explained that there is a close relation between conformal symmetry of the future boundary and the absence of an  arrow-of-time. It follows that any theory that has a time-arrow must break conformal symmetry. The tree model with terminals is an example in which we can test out the hypothesis. We will see that the existence of terminals leads to an  attractor called a fractal-flow  which breaks conformal invariance while creating a global  arrow-of-time.

The mathematical structure of a fractal-flow is describable in terms of correlation functions similar to those that we discussed earlier.
These correlation functions are defined by asking conditional questions of the following kind:

Given a point $x,$ what is the probability that it has color $n$
given that it has not been swallowed up by a terminal bubble?

Similarly, Given two points $x$ and $y,$ what is the conditional
probability that they have color $n_x$ and $n_y,$ given that neither
has been swallowed up by a terminal bubble?

Such questions define a collection of conditional correlation
functions
 \bea
 && C_n(x) \cr \cr
 && C_{n_x, n_y}(x , y) \cr \cr
 && C_{n_x, n_y, n_z}(x , y, z)\cr \cr
 &&\ldots
 \label{conditional correlators}
 \eea

The conditional correlation functions are calculated  as before with one exception. In
\ref{C=ePP} instead of the equilibrium probability we use the dominant probability
$P_n^{\{D\}}.$

The one point function $C_n(x)$ is the dominant probability,
\be
C_n(x) = P_n^{\{D\}} = e^{\frac{S_n}{2}} (D)_n  \ 2^{-\Delta_D u_0}
\label{pd}
\ee

Because the dominant probability is time-dependent,
\be
P_n^{\{D\}} \sim 2^{-\Delta_D u}
\ee
all correlations decay with time. This is a manifestation of the fact that the co-moving volume (but not proper volume) becomes filled with terminal vacua, leaving over only a fractals-worth of inflating co-moving space.
If one is only interested in relative conditional probabilities than the time-dependent factor may be dropped. Given that the point $x$ has not been swallowed by a terminal bubble, the relative probability that it has color $m$ is  $P_m^{\{D\}}.$  It should be noted that in general the dominant probability vector is not similar to the statistical weight $e^{S_m}.$ The statistical weight is dominated by the vacuum of lowest \cc \. The dominant probability is dominated by the vacuum of lowest decay rate. In general these need not be the same. More generally, the statistical weight depends only on the entropies $S_n$  and not on the decay rates. The dominant eigenvector does depend on the rates.

Next consider the two-point function. The method of calculation is the same as before, with the exception that I mentioned. Using \ref{pd}
\be
C_{nm}(x,y)= \sum_{r}
2^{-\Delta_D u_i} e^{S_{r} \over 2} \ (D)_{r}
P_{nr}(u_0 -u_i)P_{mr}(u_0 -u_i),
\label{CABD}
\ee
or substituting \ref{connector} and using
$$
2^{u_i} =\frac{1}{ |x-y|_2}
$$

we get
\be
C_{nm}(xy)= \sum_{I,J}\left\{   \sum_r e^{-S_r \over 2} \  (D)_r \  (I)_r \
(J)_r
\frac{1}{ |x-y|_2^{(\Delta_J+\Delta_I-\Delta_D)}}
% \ 2^{(\Delta_J+\Delta_I-\Delta_D ) u_i}
 \right\}
\left[
(I)_{n} \
(J)_{m}  \ e^{S_{n} +S_{m} \over 2}  \  \ 2^{-(\Delta_I+\Delta_J)u_0} \right].
\label{term 2pt}
\ee

In terms of the fields with well-defined scaling dimensions,
\be
\langle \mathcal{O}_I(x)\mathcal{O}_J(y)\rangle=   \sum_n e^{-S_n \over 2} \  (D)_n  \ (I)_n  \ (J)_n
\left({1 \over |x -y|_2}  \right)^{(\Delta_J+\Delta_I-\Delta_D ) }.
\label{insertion}
\ee

Equations \ref{term 2pt}  and \ref{insertion}  have significant differences with their equilibrium counterparts. First of all the two-point function $C_{nm}(xy)$ does not exhibit cluster decomposition. Cluster decomposition is connected to the fact that operator products typically include the unit operator with dimension zero. We saw earlier how cluster decomposition arose from the contribution of the eigenvector $(0)$ which in fact does have operator dimension $\Delta_0 = 0.$  By contrast, the theory with terminals has no eigenvector with dimension zero. The closest analog is the dominant eigenvector.  It is straightforward to calculate the contribution to \ref{term 2pt} from $(I) = (J) =(D).$ One finds a contribution proportional to
\be
\frac{1}{|x-y|_2^{\Delta_D}} P_n^D P_m^D
\ee
This is not conventional clustering which would require a term which did not go to zero with increasing $|x-y|_2.$  From this we can be sure that the conditional correlation functions do not define a conventional field theory.

From \ref{insertion} one also sees something non-standard; namely, the two point function is not diagonal in dimension. In fact comparing the two point function
 of equation \ref{insertion} with the three point function in \ref{3pt1} and \ref{3pt2}, we see that it
 closely resembles a conformal three-point function involving an insertion of the dominant field $ \mathcal{O}_D (\infty)$
evaluated at infinity. This obviously breaks conformal invariance which would require the two-point function to be diagonal.

Similar things are true for higher-point functions: there is always an extra insertion of $\mathcal{O}_D  (\infty)$  breaking conformal invariance, but the correlation functions do have definite scaling properties under scale-transformations.

This fractal-flow attractor seems to be a new type of structure quite distinct from conformal field theory. Given its connection to the arrow-of-time problem, it surely deserves more study.

\subsection{Local Fractal-Flow}

Let us return to the local viewpoint---the view from a causal patch. Earlier I showed how the equilibrium fixed-point leads to time-symmetric correlation functions characteristic of  fluctuations in thermal equilibrium. Now I want to  reconsider the issue in the theory with terminals.

As before, the first question is:
Given that a point on a causal patch has color $m$ at time $u,$ what is the probability that it will have color $n$  at the next instant, $u+1?$
$$P(n, u+1 |  m,u  ).$$

The answer is completely unchanged from the earlier discussion,
$$
P(n, u+1 |  m,u  ) = \gamma_{nm}  = M_{nm} e^{S_n}.
$$

But now consider the time-reversed question: Given that a point on a causal patch has color $m$ at time $u,$ what is the probability that it had color $n$  at the previous instant, $u-1?$ In other words what is
$$P(n, u-1 |  m,u  ).$$

  The Bayesian analysis still gives \ref{bayes}
$$
P(n, u-1 |  m,u  ) =   P(m,u  |   n, u-1   )  \frac{P(n, u-1)}{P(m,u )}.
\label{bayes'}
$$
but now the factor $ \frac{P(n, u-1)}{P(m,u )}$ is given by the dominant eigenvector. One finds
\be
P(n, u-1 |  m,u  )=  \gamma_{nm} e^{S_m -S_n} 2^{\Delta_D}\frac{P^{\{D\}}_n}{P^{\{D\}}_m}
\ee
or,
\be
\frac{P(n, u-1 |  m,u  )}{P(n, u+1 |  m,u  )}= 2^{\Delta_D}  \frac{e^{-S_n}P^{\{D\}}_n}{e^{-S_m}P^{\{D\}}_m} \ \neq 1
\ee

It is easy to generalize this to allow arbitrary (integer) time-interval $v$,
\be
\frac{P(n, u-v |  m,u  )}{P(n, u+v |  m,u  )}= 2^{(  \Delta_D) v}  \frac{e^{-S_n}P^{\{D\}}_n}{e^{-S_m}P^{\{D\}}_m} \ \ee

Thus we see that terminals induce  a  robust arrow-of-time that persists long after transients have faded, and the dominant eigenvector dominates.

\setcounter{equation}{0}
\section{Irrelevance of initial Entropy}

There is a paradoxical issue about terminals that
concerns the relevance of initial conditions for observations made within a causal patch. Let us suppose that there are $10^{500}$ vacua on the de Sitter Landscape. In this case we can expect that wherever we begin, after a few hundred transitions from one minimum to another, we will  hit a terminal. The only way to avoid doing so is for ``up-transitions" to cause repeated jumps to increased \cc. The rates for up-transitions are vastly smaller than for down transitions. Therefore from a strictly local perspective one might conclude that the probability is very high for the observed universe to be within a few hundred transitions from the initial condition.  In particular there would certainly not have been time to equilibrate  to the dominant eigenvector. If this is correct, statistical predictions about our location in the landscape crucially depend on initial conditions, about which we know very little.

On the other hand, if one looks at the tree globally, the overwhelming number of branches occur at late time. The number of branches ending at time $u_0$ grows exponentially. Therefore, if one makes the usual assumption of typicality, the best guess would be that we are extremely far from the initial condition, and from the transients associated with it.

Does  this
 matter for anything we  observe? I think it  does.  In defining a measure on the landscape, it makes a difference whether we assume a prior probability given by the dominant eigenvector, or by something else. Bousso \cite{Bousso:2009dm}, who advocates the local view, also advocates using the dominant eigenvector---his quantitative analysis of the cosmological constant depends on it. But, as Bousso admits, it is hard to see how this can be justified without some reference to the global viewpoint.

If we do take the global view then
most branches of the tree occur at such late times relative to the initial condition,  that transients will have died away. Probabilities  are then governed by the universal asymptotic  behavior and not the initial color---or entropy---of the root. If this is right,
there is no advantage to a low entropy  starting point  \cite{Bousso:2011aa}. Whatever the initial state, as long as it leads to eternal inflation the  existence of a time-arrow is guaranteed by the fractal-flow attractor.

There is an important quantitative question that I have not discussed, but which has recently addressed \cite{Bousso:2011aa}. Roughly speaking it is whether the flow is strong enough to overwhelm the tendency for Boltzmann fluctuations. One can imagine a limit in which the decay to terminals---the values of the $\gamma_n$---are vastly smaller than the other transition rates. In this limit the flow would be so feeble that it would be overwhelmed by fluctuations. Such stagnated regions of the landscape, which allow observers, are dangerous because freak  Boltzmann fluctuations in those regions could potentially outweigh all  normal environments.
\cite{Bousso:2011aa} has analyzed the quantitative question and I have nothing to add to that analysis.

\setcounter{equation}{0}
\section{The Explanatory Power of Eternal Inflation}

I want to justify  the claim that I made earlier that eternal inflation has explanatory power. I mentioned two items in addition to the arrow-of-time:
\bi
\item The small value of the cosmological constant.
\item The cosmic coincidence  that we happen to live at a time, relative to the big bang,
 comparable to the time-scale provided by the cosmological constant.
\ei
 The explanations that eternal inflation offers are statistical in nature. Here is where the large numbers of eternal inflation come in. First of all, the landscape \cite{Bousso:2004fc} \cite{Kachru:2003aw} \cite{Susskind:2003kw} must be assumed to be diverse enough  to   provide  a dense set of possible values for constants such as the cosmological constant, the parameters ordinary slow-roll inflation, and even a broad range of time-scales for biological evolution.

 Secondly, space is so big that every environment occurs many times.

 Finally time is so big that the dominant fractal-flow attractor dominates the population of vacua.

The explanatory power of eternal inflation rests to some degree on anthropic considerations which require us  to incorporate a few additional ingredients into the tree model. The new ingredients concern the existence of observers, although they do not depend on any detailed assumptions about the nature of life or intelligence. First, the usual  assumption of typicality:  the probability of an observation of a given type is proportional to the number of such observations made under the cutoff surface.  It is not our purpose to justify this assumption but only to show how the geometry and causal structure of the multiverse influence the answer.

Next, we need to define the concept of    a bubble-nucleation of a given vacuum type $n.$
 Consider an edge of type $n$ somewhere on the tree. If the rates $\gamma$ are small then the near-term causal future of that causal patch containing that edge will most likely be dominated by cells (edges) of similar type. However that is not generally true of the causal past of the edge. The causal past is defined by tracing  back along a unique series of edges of the tree.
 Eventually as one works backward, a point will occur where the color is no longer $n.$
 That point defines the nucleation event. If all rates $\gamma$ are very small, then what grows out of the nucleation point will be mostly of type $n.$
 \begin{figure}[ht]
\begin{center}
\includegraphics[scale=.3]{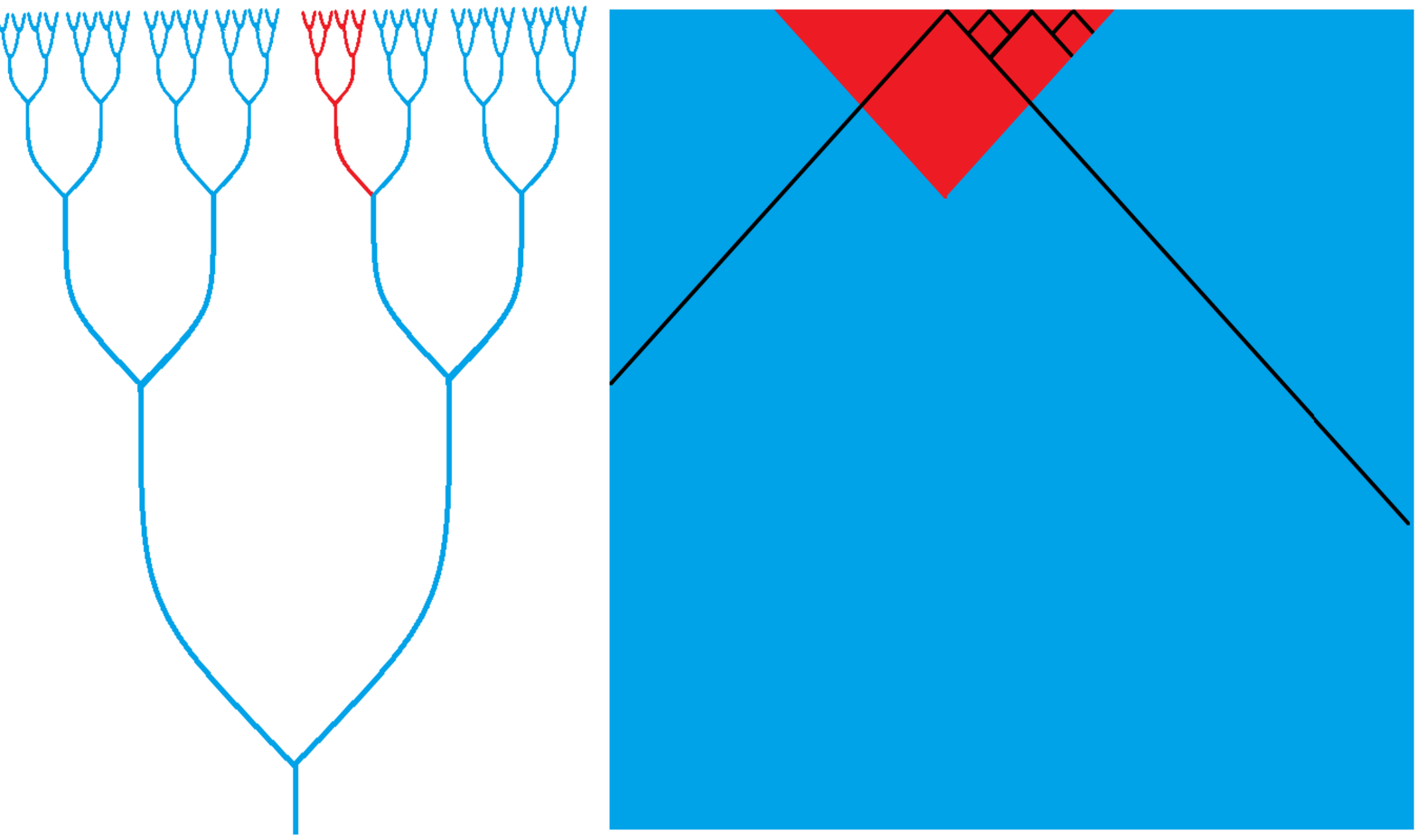}
\caption{On the right side of the figure a red bubble is shown nucleating in de Sitter space. Ignoring later nucleations, the bubble expands to fill the causal future of the nucleation point. The tree-analog of the same process is shown in the left side of the figure. }
\label{f6}
\end{center}
\end{figure}

Another  standard assumption is  that observers  in the type $n$ environment can exist  for a limited range of proper time, subsequent to the nucleation event. For simplicity we take that period to be concentrated around a proper time  $\tau_{obs}.$   This proper time can be translated into a number of links. In the $n$-vacuum the proper time of a segment composed of $L$ links is
 \be
 \tau = {L \over H_n}.
 \ee
Setting this equal to $\tau_{obs}$ gives the number of links $u_{obs}$ from the nucleation point to the place where observations take place,
\be
u_{obs} = H_n \tau_{obs}.
\ee

Let us also assume that when the nucleation of a bubble of type $n$ takes place, or shortly thereafter, an amount of matter is created within a causal patch that will eventually be assembled into a number of observers $\nu_n.$  All other things being equal, the value of $\nu_n$ will depend on $H_n.$  The form of the dependence follows from two facts: The first is that surfaces of constant proper-time $\tau$ are hyperbolic surfaces of constant negative curvature. Observers are distributed uniformly over such surfaces. The second fact is the the bounding surface of a causal patch has area $H_n^{-2}.$  Since in a hyperbolic space area and volume are proportional, it follows that the number of observers in a causal patch also varies as $H_n^{-2}.$
\be
\nu_n \sim H_n^{-2}.
\label{nu = 1/HH}
\ee

In principle we would like to count all observations that occur below the cutoff surface, but because populations exponentially increase, in an eternally inflating world it is sufficient to count observations that take place at the cutoff. Thus one defines the  the measure ${\cal{M}}(n)$ to be proportional to number of observations occurring at time $u_0$, in bubbles of type $n,$ anywhere in the multiverse. At the end of the calculation, $u_0$ is allowed to tend to infinity.  This choice of cutoff is natural in the model and in particular respects the symmetries but it is not the only possible choice.

There are several factors that go into  ${\cal{M}}(n)$. The first and most obvious is $\nu_n,$ the number of observers that form in a  single bubble of type $n.$

The next relevant consideration is the color $m$ of the immediate ancestor of the bubble. If the ancestor color is assumed to be $m$ then we must include a factor of the probability for color $m$ in the statistical ensemble. Assuming the statistical ensemble is given by the dominant eigenvector gives a factor $P^D_m.$ In addition we must include the probability $\gamma_{nm}$ that the vacuum $m$ decays to $n.$
Summing over the possible ancestor vacua leads to the factor
$$\sum_m \nu^n \ P^D_m \ \gamma_{nm}$$ or using \ref{nu = 1/HH}
\be
{\cal{M}}(n) \sim \frac{1}{H_n^2} \sum_m  \ P^D_m \ \gamma_{nm}
\ee

The final and most important factor comes from the fact that if the observations take place at time $u_0,$ the bubbles must nucleate at time $u_0 - u_{obs}.$
Consider the total number of sites available on the tree at the nucleation time.
That number is
$2^{(u_0 - u_{obs})}.$  With this factor included, the measure becomes proportional to
\be
   {\cal{M}}(n) \sim     \frac{1}{H_n^2}  \  \sum_m  \  \gamma_{nm} \ P^D_m  \  2^{(u_0 - H_n \tau_{obs})}
\ee

Now we see a possible danger: the answer appears to depend on the cutoff surface $u_0$ through the factor
$2^{u_0}$. But this factor represents the  exponential growth of all populations and should be factored out of  relative probabilities.

After dropping the cutoff dependent factor the remaining measure is
\be
   {\cal{M}}(n) \sim     \frac{1}{H_n^2}   \  \sum_m \ \gamma_{nm} \ P^D_m  \  \ 2^{- H_n \tau_{obs}}.
\ee
The factors that interests us most is the
 ones containing the hubble constant $H_n,$
\be
{\cal{M}}(n) \sim         \frac{1}{H_n^2}  \  2^{ - H_n \tau_{obs}}
\label{cc measure}
\ee

We may interpret this measure as the probability density for observations of a cosmological constant
proportional to $H_n^2. $
\be
\frac{dP(\lambda, \tau_{obs})}{d\lambda} \sim \frac{1}{\lambda} 2^{ - H_n \tau_{obs}}
\label{final measure}
\ee

The fact that the base of the exponential is $2$ and not some other number is an artifact of the
fact that the tree branches into to edges at each node. In a more realistic theory the base would be
$e^3$ corresponding to the growth of volume $e^{3H}$ during each e-folding. With that substitution \ref{final measure}  agrees exactly with the causal-patch and light-cone measure of Bousso and collaborators, and the following conclusions can be taken straight from \cite{Bousso:2010im}. In particular the \ref{final measure} can be interpreted as a joint probability for the cosmological constant $\lambda$ and the observer-time $\tau_{obs}.$  The factor $ \frac{1}{\lambda}$  favors the smallest values of $\lambda$ that are consistent with the existence of observers. The exponential factor may be thought of as a probability measure of the time of observation. As such, it favors $\tau_{obs} \sim H^{-1},$ consistent with the cosmic coincidence observation.

\section*{Acknowledgements}

My interest in the arrow-of-time dates back to discussions with Savas Dimopoulos in 1978. We were working on baryogenesis  and together, realized that time-asymmetry was one of the necessary conditions for baryon-asymmetry. Although he may not remember it all, I am grateful to Savas for many insights into the nature of time-asymmetry.

The material in this lecture represents work done in collaboration with Daniel Harlow, Douglas Stanford, and Stephen Shenker (It should not be referenced without also referencing \cite{Harlow:2011az}). Much of what is in this lecture is due to them.

 I have also benefited from stimulating discussions with Raphael Bousso.

This work is supported by the Stanford Institute for Theoretical Physics and NSF Grant 0756174.

\end{document}